\documentstyle[aps,epsf]{revtex}

\newcommand{\beq}{\begin{equation}}
\newcommand{\eeq}{\end{equation}}
\newcommand{\bear}{\begin{eqnarray}}
\newcommand{\ear}{\end{eqnarray}}
\newcommand{\earn}{\nonumber \end{eqnarray}}
\newcommand{\dst}{\displaystyle}

\newcommand{\phisq}{\langle\phi^2\rangle}
\newcommand{\Tmn}{\langle T_{\mu\nu}\rangle}
\newcommand{\rc}{r_{\rm c}}
\newcommand{\rg}{r_{\rm g}}

\arraycolsep=\mathsurround  

\begin{document}
 
\title{Analitical approximation of $\langle\phi^2\rangle$ for a
massive scalar field in static spherically symmetric spacetimes}

\author{Sergey V. Sushkov\thanks{e-mail: sushkov@kspu.kcn.ru}}  

\address{Department of Geometry, Kazan State Pedagogical University,\\
Mezhlauk 1 st., Kazan 420021, Russia}

\maketitle

\begin{abstract}
An analitical approximation of $\langle\phi^2\rangle$ for a massive scalar
field in a zero temperature vacuum state in static spherically symmetric
spacetimes is obtained. The calculations are based on the method for
computing vacuum expectations values for scalar fields in general static
spherically symmetric spacetimes derived by Anderson, Hiscock and Samuel
\cite{Anderson,AHS}. The analitical approximation is used to compute
$\phisq$ in Schwarzschild and wormhole spacetimes.

\vskip12pt\noindent
{\normalsize PACS number(s): 04.62.+v, 04.70.Dy}
\end{abstract}

\section{Introduction}
In the absence of a fully satisfactory theory of quantum gravity the
semiclassical theory of gravity takes an important role in studies of
effects of the back reaction of the quantized fields upon the spacetime
geometry. A primary technical difficulty in semiclassical gravity is that
expectation values have the strong functional dependence on the metric
tensor $g_{\mu\nu}$ and are generally impossible to be calculated
analitically. For this reason, much efforts have been concentrated on
developing approximate methods.

In 1982, Page \cite{Page} developed a method for obtaining an approximate
expression for $\langle T_{\mu\nu} \rangle$ for a conformal massless scalar
field in static Einstein spaces. Later a slightly different approach was
proposed by Brown and Ottewill \cite{BO} (see also Ref. \cite{BOP}). In 1987,
Frolov and Zel'nikov \cite{FZ} constructed the approximate expression for
$\langle T_{\mu\nu} \rangle$ for conformal massless fields in general static
spacetimes; their construction was based on pure geometrical arguments and
common properties of the stress-energy tensor.  The effective action approach
had been elaborated by a number of authors
\cite{FZ1,FZ2,FZ3,FZ4,FSZ,Avra1,Avra2,Avra3,Mat}.
Recently, Anderson \cite{Anderson} and Anderson, Hiscock and Samuel
\cite{AHS} developed a method of deriving an approximation for $\phisq$
and $\Tmn$ in a static spherically symmetric spacetime using the WKB
approximation for the modes of the scalar field; the method is suitable
for the fields with an arbitrary coupling $\xi$ to the scalar curvature;
also the fields can be either in a zero temperature vacuum state or a
nonzero temperature thermal state.

The aim of this paper is to obtain the approximate expression for $\phisq$
using the Anderson-Hiskock-Samuel approach. Note that a practical applying of
the method faces with considerable technical difficulty. It consists in
necessity of computing the mode sums and then expanding them in powers of
$\epsilon$ (where $\epsilon=i(t-t')$ characterizes the point splitting). In
Refs. \cite{Anderson,AHS} such calculations was made approximately in the
large $\omega$ limit. In this paper we present exact results on computing the
mode sums. 

Another difficulty has no computational nature. The problem is that a
zeroth-order solution, which determines an iterative procedure of solving the
mode equation, can be only defined up to terms of the second order. Of
course, in principle, a final result of the iterative procedure should not
depend on a particular choice of the zeroth-order solution. However, the form
of an approximate solution of the second or higher order should depend on
this choice. In this paper we solve the mode equation iteratively using the
zeroth-order solution which is slightly different from that in Refs.
\cite{Anderson,AHS} used.
  
The paper is organized as follows. In Sec. II, following the
Anderson-Hiscock-Samuel approach, we develop the second-order WKB
approximation for an unrenormalized expression for $\phisq$, where $\phi$ is
a massive scalar field with arbitrary curvature coupling in a general static
spherically symmetric spacetime. In Sec. III the resulting expression is
renormalized using the method of covariant point splitting. In Secs. IV and V
we apply the resulting analitical approximation for $\phisq$ to investigate
the vacuum polarization of a massive scalar field in a Schwarzschild
spacetime and in a wormhole spacetime, respectively. In Appendix A we derive
the asymptotical expansions for the mode sums and integrals in the
unrenormalized expression for $\phi$. In Appendix B some technical
details are discussed. 

The units $\hbar=c=G=1$ are used throughout the paper. 

\section{An unrenormalized second-order WKB approximation for $\phisq$} 
In this section we derive an unrenormalized expression for $\phisq$
for a massive scalar field in an arbitrary static spherically symmetric
spacetime. Note that at this stage our consideration will mainly follow
the papers by Anderson \cite{Anderson} and Anderson, Hiscock, and Samuel
\cite{AHS} in which one may find some additional details.

As in Refs. \cite{Anderson,AHS}, a Euclidean space approach is used. The
metric for a general static spherically symmetric spacetime when continued
analitically into Euclidean space is
        \beq\label{metric}
        ds^2=f(r)d\tau^2+h(r)dr^2+r^2(d\theta^2+\sin^2\theta\, d\varphi^2). 
        \eeq
Here $\tau=it$ is the Euclidean time, and $f$ and $h$ are
arbitrary functions of $r$ which, if the space is asymptotically
flat, become constant in the limit $r\to\infty$. 

Consider a quantized scalar field $\phi$ with mass $m$ and coupling $\xi$
to the scalar curvature $R$. We assume that the field is in a vacuum 
state defined with respect to the Killing vector which always
exists in a static spacetime. An unrenormalized expression for $\phisq$
can be computed from the Euclidean Green's function $G_E(x,\tilde x)$. This
expression given in Refs. \cite{Anderson,AHS} is
        \bear \label{phisqunren}
        \langle\phi^2(x)\rangle_{\rm unren}&=&
        G_E(x,\tau;x,\tilde\tau) 
        \nonumber \\
        &=&\frac1{4\pi^2}\int_0^\infty d\omega 
        \cos[\omega(\tau-\tilde\tau)] 
        \sum_{l=0}^\infty \left[(2l+1)p_{\omega l}q_{\omega l}
        -\frac1{rf^{1/2}} \right],
        \ear
where the modes $p_{\omega l}$ and $q_{\omega l}$ obey the equation
        \bear\label{modeeqn}
        \frac1h \frac{d^2S}{dr^2}+\left[\frac2{rh}\right.&+&
        \left.\frac1{2fh}\frac{df}{dr}-\frac1{2h^2}\frac{dh}{dr}
        \right]\frac{dS}{dr}
        -\left[\frac{\omega^2}{f}+\frac{l(l+1)}{r^2}+\xi R\right]S=0.
        \ear 
They also satisfy the Wronskian condition
        \beq\label{wronskian}
        C_{\omega l}\left[p_{\omega l}\frac{dq_{\omega l}}{dr}-
        q_{\omega l}\frac{dp_{\omega l}}{dr}\right]=-\frac1{r^2}
        \left(\frac hf \right)^{1/2},
        \eeq
where $C_{\omega l}$ is a normalization constant. 

Let us stress that points in Eq. (\ref{phisqunren}) are splitted. Namely,
points are separated in time so that $\epsilon\equiv(\tau-\tilde\tau)$,
$\tilde r=r$, $\tilde \theta=\theta$, $\tilde \varphi=\varphi$. As was
first pointed out by Candelas and Howard \cite{CH} for the case of
Schwarzschild spacetime, the Euclidean Green's function have superficial
divergences with this separation of points. As discussed in Refs.
\cite{Anderson,AHS}, these cannot be real divergences because the Green's
function must be finite when the points are separated; to remove the
divergences one has to subtract some additional counterterms. The terms
$(rf^{1/2})^{-1}$ in brackets in Eq. (\ref{phisqunren}) are such
counterterms (for details, see discussion in Refs. \cite{Anderson,AHS}). 

There is a WKB representation for the modes which is very useful
in calculations of $\phisq$. The WKB representation is obtained
by the change of variables
        \beq\label{modes}
        \begin{array}{l}
        \displaystyle
        p_{\omega l}=\frac1{(2 r^2 W)^{1/2}}
        \exp\left[\int^r W \left(\frac hf\right)^{1/2} dr\right], \\ 
        \\
        \displaystyle
        q_{\omega l}=\frac1{(2 r^2 W)^{1/2}}
        \exp\left\{-\left[\int^r W \left(\frac hf\right)^{1/2} 
        dr\right]\right\},
        \end{array}
        \eeq
where $W$ is a new function of $r$.
Substitution of Eq. (\ref{modes}) into Eq. (\ref{wronskian}) shows that
the Wronskian condition is obeyed if $C_{\omega l}=1$, and substitution of
Eq. (\ref{modes}) into Eq. (\ref{modeeqn}) gives the following equation
for $W$: 
        \bear\label{eqn4W}
        W^2&=&\omega^2+m^2 f+l(l+1)\frac{f}{r^2}
        +\frac{1}{2r}\left(\frac{f}{h}\right)'
        + \xi R f+\frac12\frac{f}{h}\frac{W''}{W}
        +\frac14\left(\frac fh\right)'\frac{W'}{W}
        -\frac34\frac{f}{h}\frac{W'^2}{W^2},
        \ear
where the scalar curvature $R$ is
        $$
        R=-\frac{f''}{fh}
        +\frac{f'^2}{2f^2h}+\frac{f'h'}{2fh^2}
        -\frac{2f'}{rfh}+\frac{2h'}{rh^2}
        -\frac2{r^2h}+\frac2{r^2},
        $$
and the prime denotes the derivative with respect to $r$.
The usual way to work out Eq. (\ref{eqn4W}) is to solve it iteratively.
It is obvious that a choice of zeroth-order iteration determines the further
iterative procedure. In Refs. \cite{Anderson,AHS} it was suggested to take 
        $$
        \overline W^{(0)}=\overline\Omega(r)=
        \left[\omega^2+m^2 f+(l+1/2)^2\frac{f}{r^2}\right]^{1/2}
        $$  
as the zeroth-order solution. This choice seems to be convenient because
it simplifies some further calculations. However, this is not the best
choice. Really, consider the asymptotically flat region of spacetime
where $f,h\to 1$ and $f',f'',h',h''\to 0$. Assume also that $W',W''\to 0$
there. In this case the equation (\ref{eqn4W}) reduces to 
        $$
        W^2=\Omega^2(r)
        $$  
with 
        \beq
        \Omega(r)=\left[\omega^2+m^2 f+l(l+1)\frac{f}{r^2}\right]^{1/2}.
        \eeq
Hence, it is seen that a natural choice for the zeroth-order solution
is $W^{(0)}=\Omega$.  Note that $W^{(0)}=\overline W^{(0)}-f(4r^2)^{-1}$.

So, we take $W^{(0)}=\Omega$ as the zeroth-order solution. Then, the
second-order solution of Eq. (\ref{eqn4W}) is
        \bear\label{sos}
        W&=&\Omega+W^{(2)} 
        =\Omega+\frac{V_1+V_2}{2\Omega} 
        +\frac14\frac{f}{h}\frac{\Omega''}{\Omega^2}
        +\frac18\left(\frac{f}{h}\right)'\frac{\Omega'}{\Omega^2}
        -\frac38\frac{f}{h}\frac{\Omega'^2}{\Omega^3}, 
        \ear
with
        $$
        V_1(r)=\frac{1}{2r}\left(\frac{f}{h}\right)', \quad
        V_2(r)=\xi R f. 
        $$

Discuss the question about applicability of the second-order solution.
It is correct provided $\Omega>\!\!>|W^{(2)}|$. This condition must be
fulfilled for each mode. In particular, for the lowest null mode with
$\omega=0$ and $l=0$ we obtain
        \bear\label{inequal}
        m^2&>\!\!>&\left|\frac12 \xi R+\frac14\frac{f'}{fhr}-
        \frac14\frac{h'}{h^2r}
        +\frac18\frac{f''}{fh}\right.
        \left.
        -\frac{1}{16}\frac{f'h'}{fh^2}
        -\frac{3}{32}\frac{f'^2}{f^2h}\right|.
        \ear 
In the other words, the last condition means that the Compton length
$\rc=m^{-1}$ is much smaller than the characteristic length scale of the
spacetime geometry. Restrictions, which the inequality (\ref{inequal})
imposes for the metric functions $f$ and $h$, ensure correctness of the
WKB approximation. Hereinafter we shall suppose that $f$ and $h$ answer
these requirements. 

To obtain the second-order WKB approximation for $\phisq$ we substitute
Eq. (\ref{sos}) into (\ref{modes}), and then into (\ref{phisqunren}).
Neglecting terms of the fourth order and higher we can finally find
        \bear\label{phisqwkbunren} 
        \phisq_{\rm unren}= && \frac1{4\pi^2}
        \int_0^\infty d\omega\cos[\omega(\tau-\tau')] 
        \sum_{l=0}^\infty 
        \left\{\left(\frac{l+1/2}{r^2\Omega}-\frac1{r\sqrt{f}}\right)
        -\frac{l+1/2}{r^2\Omega^3}  \frac{V_1+V_2}{2} 
        \right.
        \nonumber \\
        && 
        -\frac18\frac{l+1/2}{r^2\Omega^5}
        \left[
        \frac{f}{h}U''+\frac12\left(\frac fh\right)'U'\right]
        \left.
        +\frac5{32}\frac{l+1/2}{r^2\Omega^7}\frac fh U'^2
        \right\},
        \ear
with
        $$
        U(r)=m^2 f+l(l+1)\frac{f}{r^2}.
        $$

\section{A renormalized expression for $\phisq$} 
In this section we derive a renormalized expression for $\phisq$ in the
framework of the second-order WKB approximation.

As is usual for the method of point splitting, in order to renormalize
$\phisq_{\rm unren}$ one should subtruct renormalization counterterms 
from $\phisq_{\rm unren}$ and take the limit $\epsilon \to 0$
($\tau'\to\tau$). Schematically,
        \beq\label{subtr}
        \phisq_{\rm ren}=\lim_{\epsilon\to 0}
        \left(\phisq_{\rm unren}-\phisq_{\rm DS}\right).
        \eeq
The renormalization counterterms for a scalar field has been
obtained by Christensen \cite{Chris} which used the DeWitt-Schwinger
expansion for the Feynman Green's function. In the case of a massive
scalar field they are given by 
        \bear
        \phisq_{\rm DS} &=& G_{\rm DS}(x,x') \nonumber \\
        &=&\frac1{8\pi^2\sigma}+\frac1{8\pi^2}
        \left[m^2+\left(\xi-\frac16\right)R\right]
        \left[C+\frac12\ln\left( \frac{m^2|\sigma|}{2}
        \right)\right]  
        -\frac{m^2}{16\pi^2}+\frac1{96\pi^2}R_{\alpha\beta}
        \frac{\sigma^\alpha\sigma^\beta}{\sigma}.
        \ear
Here $\sigma$ is equal to one half the square of the distance between the
points $x$ and $x'$ along the shortest geodesic connecting them. $C$ is
Euler's constant, $R_{\alpha\beta}$ is the Ricci tensor and
$\sigma^\alpha\equiv\sigma^{;\alpha}$. 

To perform the procedure of renormalization in practice, one should expand
the expressions for $\phisq_{\rm unren}$ and $\phisq_{\rm DS}$ in powers
of $\epsilon$. The expansion of $\phisq_{\rm DS}$ can be easily found by
using the expansion of $\sigma$ and its derivatives in powers of
$\epsilon$ given in Refs. \cite{Anderson,AHS}:
        \beq
        \begin{array}{l}
        \dst
        \sigma=\frac12 f\epsilon^2+{\rm O}(\epsilon^4),\quad 
        \sigma^\tau=-\epsilon+\frac{f'^2}{24fh}\epsilon^3
        +{\rm O}(\epsilon^5), \\[12pt]
        \dst
        \sigma^r=\frac{f'}{4h}\epsilon^2+{\rm O}(\epsilon^4), \quad
        \sigma^\theta=\sigma^\phi=0.
        \end{array}
        \eeq
In general, we have 
$\phisq_{\rm DS}=a\epsilon^{-2}+b\ln\epsilon+c+O(\epsilon^2)$ 
where first two terms describe ultraviolet divergences. It is well-known
that the second-order WKB approximation for $\phisq_{\rm unren}$ contains
the same divergences, and so the subtraction of $\phisq_{\rm DS}$
annihilates terms of $\phisq_{\rm unren}$ that is infinite in the limit
$\epsilon\to 0$. 

To obtain the expansion of $\phisq_{\rm unren}$ in powers of
$\epsilon$ we rewrite Eq. (\ref{phisqwkbunren}) in the form
being more convenient for further analysis:
        \bear\label{phisqconv}
        4\pi^2\phisq_{\rm unren}&=&\frac1{r^2}S_0(\epsilon,\mu)-
        S_1(\epsilon,\mu)\frac{V_1+V_2}{2f}
        -S_2(\epsilon,\mu)\frac{r^2}{8f^2}
        \left[
        \frac{f}{h}\left(\frac{f}{r^2}\right)''
        +\frac12\left(\frac fh\right)'\left(\frac{f}{r^2}\right)'
        \right]
        \nonumber \\
        &&
        +S_3(\epsilon,\mu)\frac{5r^4}{32f^2h} 
        \left(\frac{f}{r^2}\right)'^2 
        -N_2^1(\epsilon,\mu)\frac{r^2}{8f^2}
        \left[
        \frac{f}{h}\left(\frac{f}{r^2}\mu^2\right)''
        +\frac12\left(\frac fh\right)'\left(\frac{f}{r^2}\mu^2\right)'
        \right]
        \nonumber \\
        &&
        +N_3^1(\epsilon,\mu)\frac{5r^4}{32f^2h}\left(\frac{f}{r^2}\mu^2\right)'^2
        +N_3^3(\epsilon,\mu)\frac{5r^4}{16f^2h}\left(\frac{f}{r^2}\right)'
        \left(\frac{f}{r^2}\mu^2\right)'
        \ear
with 

\renewcommand{\theequation}{\arabic{equation}.1}
        \bear\label{s0}
        S_0(\epsilon,\mu)&=&\int_0^\infty du 
        \cos\left(u\frac{\epsilon\sqrt{f}}{r}\right)
        \sum_{l=0}^\infty
        \left[\frac{l+1/2}{\sqrt{u^2+\mu^2+(l+1/2)^2}}-1\right],
        \ear

\addtocounter{equation}{-1}
\renewcommand{\theequation}{\arabic{equation}.2}
        \bear\label{sn}
        S_n(\epsilon,\mu)&=&\int_0^\infty du 
        \cos\left(u\frac{\epsilon\sqrt{f}}{r}\right)
        \sum_{l=0}^\infty
        \frac{(l+1/2)^{2n-1}}{[u^2+\mu^2+(l+1/2)^2]^{n+1/2}},
        \\
        \lefteqn{n=1,2,3,\dots,}\phantom{S_n(\epsilon,\mu)}
        \nonumber
        \ear

\addtocounter{equation}{-1}
\renewcommand{\theequation}{\arabic{equation}.3}
        \bear
        N_n^m(\epsilon,\mu)&=&\int_0^\infty du 
        \cos\left(u\frac{\epsilon\sqrt{f}}{r}\right)
        \sum_{l=0}^\infty
        \frac{(l+1/2)^m}{[u^2+\mu^2+(l+1/2)^2]^{n+1/2}},
        \\
        \lefteqn{n=1,2,3,\dots,}\phantom{N_n^m(\epsilon,\mu)} 
        \nonumber \\ 
        \lefteqn{m=1,3,\dots,2n-3}\phantom{N_n^m(\epsilon,\mu)} 
        \nonumber
        \ear
where we denote
\renewcommand{\theequation}{\arabic{equation}}
        \beq
        \mu^2\equiv m^2 r^2 -\frac14.
        \eeq
Note that $\mu^2>0$ if $r>(2m)^{-1}$. 

Now one must calculate the asymptotical expansion for the functions 
$S_0(\epsilon,\mu)$, $S_n(\epsilon,\mu)$ and $N_n^m(\epsilon,\mu)$ in the
limit $\epsilon\to 0$. In Appendix A details of such calculations are
given. Here we present some final formulas: 

        \bear \label{S}
        S_0(\epsilon,\mu)&=&\frac{r^2}{f\epsilon}
        +\frac12\left(\mu^2-\frac{1}{12}\right)
        \left(\ln\frac{\epsilon\mu}{2}+C\right) 
        -\frac{\mu^2}{2}-\mu^2 S_0(2\pi\mu)+{\rm O}(\epsilon^2\ln\epsilon),
        \\[6pt]
        S_n(\epsilon,\mu)&=&
        -\frac{2^{n-1}(n-1)!}{(2n-1)!!}
        \left(\ln\frac{\epsilon\mu}{2}+C\right)
        +\frac{S_n(2\pi\mu)}{(2n-1)!!}+{\rm O}(\epsilon^2\ln\epsilon),
        \nonumber
        \ear
where $(2n-1)!!\equiv 1\cdot3\cdot5\cdots(2n-1)$, $n=1,2,3,\dots$ and
$S_0(2\pi\mu)$ and $S_n(2\pi\mu)$ are determined by Eqs. (\ref{appS}).

Note that the functions $N_n^m(\epsilon,\mu)$ are regular in the limit
$\epsilon\to 0$. Hence one may directly put $\epsilon=0$, so that
        \beq\label{Nnm}
        N_m^m(0,\mu)=\int_0^\infty du\, \sum_{l=0}^\infty
        \frac{(l+\frac12)^m}{[u^2+\mu^2+(l+\frac12)^2]^{n+1/2}}.
        \eeq
Changing the order of summation and integration in Eq. (\ref{Nnm}) and
integrating over $u$ gives, in particular,

\renewcommand{\theequation}{\arabic{equation}.1}
        \beq\label{N}
        N_2^1(0,\mu)\equiv\frac23N_2^1(\mu)=\frac23\sum_{l=0}^\infty
        \frac{l+\frac12}{[\mu^2+(l+\frac12)^2]^2},
        \eeq
\addtocounter{equation}{-1}
\renewcommand{\theequation}{\arabic{equation}.2}
        \bear
        &&N_3^i(0,\mu)\equiv\frac{8}{15}N_3^i(\mu)=
        \frac{8}{15}\sum_{l=0}^\infty
        \frac{(l+\frac12)^i}{[\mu^2+(l+\frac12)^2]^3},
        \ear
\renewcommand{\theequation}{\arabic{equation}}
where $i=1,3$.

Note also that the functions $S_n(2\pi\mu)$ and $N_n^m(\mu)$ have a simple
asymptotical form at large values of $\mu$, or, correspondingly, at large
values of $mr$ (see Appendix B for details):
        \bear\label{asymSN}
        &&S_0(2\pi\mu)\approx-\frac{7}{1920(mr)^4}, \quad
        S_1(2\pi\mu)\approx\frac{1}{24(mr)^2}, \quad
        \nonumber \\
        &&
        S_2(2\pi\mu)\approx-\frac{0.0146}{(mr)^4}, \quad
        S_3(2\pi\mu)\approx\frac{0.03}{(mr)^6},
        \nonumber \\
        &&
        N_2^1(\mu)\approx\frac{1}{2(mr)^2}, \quad
        N_3^1(\mu)\approx\frac{1}{4(mr)^4}, \quad
        \\
        &&
        N_3^3(\mu)\approx\frac{1}{4(mr)^2}.
        \nonumber 
        \ear

Substituting Eqs. (\ref{S},\ref{N}) into Eq. (\ref{phisqconv}) we may
find the asymptotical expansion for $\phisq_{\rm unren}$ in the limit
$\epsilon\to 0$. Finally, carring out the procedure of renormalization
(\ref{subtr}), i.e. subtracting $\phisq_{\rm DS}$, we obtain the
renormalized expression for $\phisq$ in the framework of the second-order
WKB approximation: 
        \bear\label{phisqren}
        4\pi^2\phisq_{\rm ren}=&&
        \frac14\left[m^2+(\xi-\frac16)R\right]
        \ln\left(1-\frac{1}{4m^2r^2}\right)
        +\frac{1}{16r^2}
        +\frac{f''}{24fh}-\frac{f'^2}{24f^2}
        -\frac{f'h'}{48fh^2}+\frac{f'}{12fhr}
        \nonumber \\
        &&
        -m^2 S_0(2\pi\mu)\left(1-\frac{1}{4m^2r^2}\right)
        -S_1(2\pi\mu)\frac{V_1+V_2}{2f}
        -S_2(2\pi\mu)\frac{r^2}{24f^2}
        \left[
        \frac{f}{h}\left(\frac{f}{r^2}\right)''
        +\frac12\left(\frac fh\right)'\left(\frac{f}{r^2}\right)'
        \right]
        \nonumber \\
        &&
        +S_3(2\pi\mu)\frac{r^4}{96f^2h} 
        \left(\frac{f}{r^2}\right)'^2 
        -N_2^1(\mu)\frac{r^2}{8f^2}
        \left[
        \frac{f}{h}\left(\frac{f}{r^2}\mu^2\right)''
        +\frac12\left(\frac fh\right)'\left(\frac{f}{r^2}\mu^2\right)'
        \right]
        \nonumber \\
        &&
        +N_3^1(\mu)\frac{5r^4}{32f^2h}\left(\frac{f}{r^2}\mu^2\right)'^2
        +N_3^3(\mu)\frac{5r^4}{16f^2h}\left(\frac{f}{r^2}\right)'
        \left(\frac{f}{r^2}\mu^2\right)'.
        \ear

\section{$\phisq$ in a Schwarzschild spacetime}
In this section we apply the analitical approximation for $\phisq$,
obtained above, to investigate the vacuum polarization of a massive scalar
field with arbitrary curvature coupling in a Schwarzschild spacetime.

For a Schwarzschild spacetime the metric functions $f$ and $h$ are
        \beq
        f=h^{-1}=1-\frac{2M}{r},
        \eeq
where $M$ is the mass of the black hole. The event horizon is at 
        \beq
        \rg=2M.
        \eeq
The scalar curvature $R$ is identically zero for a Schwarzschild
spacetime. 

We shall compute $\phisq$ in the region exterior to the event
horizon where the spacetime is static, i.e. $r>\rg$.
The condition (\ref{inequal}), ensuring applicability of the WKB
approximation in this region, can be rewritten as
        \beq\label{inequalins}
        \left(\frac{\rg}{\rc}\right)^2>\!\!>
        \left|\frac1{4\rho^3}-\frac{1}{32\rho^4(1-\rho^{-1})}\right|,
        \eeq
where $\rc=m^{-1}$ is Compton radius corresponding to a scalar field with
the mass $m$, and $\rho=r/\rg$ is the dimensionless coordinate, $\rho>1$.
For example, let us take $\rg=2M_{\odot}=1.48\cdot 10^{5}{\rm cm}$ and
$\rc=m_{\rm e}^{-1}=3.86\cdot 10^{-11}{\rm cm}$ where $M_{\odot}$ is the
Sun mass and $m_{\rm e}$ is the electron mass. In this case the inequality
(\ref{inequalins}) is fulfilled for $\rho-1>10^{-30}$. 

We shall assume that $\rg/\rc>\!\!> 1$, and so $mr=(\rg/\rc)\rho>\!\!> 1$
because of $\rho>1$. Therefore, the functions $S_n$ and $N_n^m$ can be
taken in their asymptotical form (\ref{asymSN}) and the expression
(\ref{phisqren}) for $\phisq$ can be written down as follows: 
        \bear\label{phisqinbh}
        16\pi^2 M^2 \phisq_{\rm ren}= &&
        \frac14\Lambda^2\ln\left(1-\frac1{4\Lambda^2\rho^2}\right)
        +\frac1{16\rho^2}
        +\frac7{1920}\frac1{\Lambda^2\rho^4}
        -\frac1{48}\frac1{\Lambda^2\rho^5}
        -\frac7{7680}\frac1{\Lambda^4\rho^6}
        \nonumber \\
        &&
        -\frac{a_1}{\Lambda^4\rho^6(1-\rho^{-1})^2}
        \left(1-\frac{13}{3\rho}+\frac{35}{6\rho^2}-\frac{5}{2\rho^3}\right) 
        +\frac{a_2}{\Lambda^6\rho^8(1-\rho^{-1})}
        \left(1-\frac3\rho+\frac{9}{4\rho^2}\right),
        \ear
where $\Lambda\equiv 2Mm=\rg/\rc$, $a_1=3,646\cdot 10^{-3}$ and 
$a_2=1.281\cdot 10^{-3}$. 
Note also that at large values of $\rho$, i.e. far from the event horizon,
the last expression has a simple asymptotical form. Taking into account
that $\ln(1-\frac1x)\approx -(\frac1x+\frac1{2x^2})$ for large
values of $x$ we obtain 
        \beq
        16\pi^2 M^2 \phisq_{\rm ren} \approx 
        -\frac1{240\Lambda^2\rho^4}.
        \eeq
Some results of numerical calculations using the analitical
approximation (\ref{phisqinbh}) are shown in Fig. 1.

\section{$\phisq$ in a wormhole spacetime}

In this section we apply the analitical approximation (\ref{phisqren}) to
calculate $\phisq$ for a massive scalar field in a wormhole spacetime.

Here we shall regard a wormhole as a time independent, nonrotating, and
spherically symmetric bridge connecting two asymptotically flat regions.
The metric of wormhole spacetime (continued analitically into Euclidean
space) can be taken in the form, suggested by 
Morris and Thorne \cite{MT}:
        \beq\label{whmetric}
        ds^2=f(l)d\tau^2+dl^2+r^2(l)(d\theta^2+\sin^2\theta\, d\varphi^2), 
        \eeq
where $l$ is the proper radial distance, $l\in(-\infty,+\infty)$. We
assume that the redshift function $f(l)$ is everywhere finite (no event
horizons); the shape function $r(l)$ has the global minimum at $l=0$, so
that $r_0=\min\{r(l)\}=r(0)$ is the radius of the wormhole throat. In
order for the spacetime geometry to tend to an appropriate asymptotically
flat limit at $l\to\pm\infty$ we impose 
        \bear
        \lim_{l\to\pm\infty}\{r(l)/|l|\}=1, \quad {\rm and} \quad
        \lim_{l\to\pm\infty}f(l)=f_{\pm}.
        \ear
For simplicity we also assume symmetry under interchange of the two
asymptotically flat regions, $l\leftrightarrow -l$, that is, 
$r(l)=r(-l)$ and $f(l)=f(-l)$. 

Note that two metrics, (\ref{metric}) and (\ref{whmetric}), can be
derived from each other by substitution 
        \beq
        \frac{dr}{dl}=\pm\frac1{\sqrt{h}}.
        \eeq
Now 
        \beq
        \frac 1h=r'^2, \quad \frac 1h \frac{dh}{dr}=r'',
        \eeq
where the prime denotes the derivative with respect to $l$.

Let $\phi$ be a massive scalar field with minimal curvature coupling
($\xi=0$) in the wormhole spacetime. Consider the simplest example, given
in Ref. \cite{MT}, when the metric functions $f(l)$ and $r(l)$ are
        \beq
        f(l)\equiv1,\quad r(l)=\sqrt{l^2+r_0^2}.
        \eeq
The condition (\ref{inequal}), ensuring applicability of the WKB
approximation in the wormhole spacetime, reduces in this case to
        \beq
        m^2>\!\!>\frac12\frac{r''}{r}=\frac12\frac{r_0^2}{(l^2+r_0^2)}.
        \eeq
It is seen that this inequality is satisfied for all values of $l$
provided $r_0^2>\!\!>(2m)^{-1}=\rc/2$. This means that the throat's radius
has to be much greater than Compton radius. If so, then $mr>\!\!>1$, and
the functions $S_n$ and $N_n^m$ can be taken in their asymptotical form
(\ref{asymSN}). Now the analitical approximation (\ref{phisqren}) for
$\phisq$ can be written as
        \bear\label{phisqinwh}
        16\pi^2\rc^2  \phisq_{\rm ren}= &&
        \left[1+\frac13\frac{\rho_0^2}{(\rho^2+\rho_0^2)^2}\right]
        \ln\left(1-\frac14\frac{1}{\rho^2+\rho_0^2}\right)
        +\frac14\frac1{\rho^2+\rho_0^2}-\frac{b_1}{(\rho^2+\rho_0^2)^2}
        +\frac{b_2}{(\rho^2+\rho_0^2)^3}
        \nonumber \\
        &&
        +\frac{\rho^2}{(\rho^2+\rho_0^2)^3}
        \left[b_3+\frac{b_4}{\rho^2+\rho_0^2}
        +\frac{b_5}{(\rho^2+\rho_0^2)^2}\right],
        \ear
where $\rho=ml$ is the dimensionless proper radial distance, 
$\rho_0=mr_0$ is the dimensionless throat's radius, and the numbers $b_i$
are $b_1=\frac{31}{160}$, $b_2=8.51\cdot10^{-3}$, $b_3=\frac{13}{48}$, 
$b_4=5.851\cdot10^{-2}$, $b_5=5.126\cdot10^{-3}$.
Note also that far from the wormhole's throat, $\rho\to\infty$, the
analitical approximation (\ref{phisqinwh}) has the simple asymptotical
form: 
        \beq
        16\pi^2 \rc^2 \phisq_{\rm ren} \approx 
        \frac{11}{240\rho^4},
        \eeq
which does not depend on the throat's radius.

In Fig. 2 we present some results of numerical calculations obtained by
using the analitical approximation (\ref{phisqinwh}).

\section*{Acknowledgments}
This work was supported by the Russian Foundation for Basic
Research grant No 99-02-17941 and the ISSEP grant No d99-263.

\section*{Appendix A: Asymptotical expansion for 
$S_{n}(\epsilon,\mu)$ in the limit $\epsilon\to 0$} 

\setcounter{equation}{0} 
\renewcommand{\theequation}{A.\arabic{equation}}
In this appendix we derive the asymptotical expansion for the functions
$S_0(\epsilon,\mu)$ and $S_n(\epsilon,\mu)$ in the limit $\epsilon\to 0$. 

Consider the following sums over $l$:
        \bear \label{sum0}
        s_0(y)=\sum_{l=0}^\infty 
        \left[\frac{l+1/2}{\sqrt{y^2+(l+1/2)^2}}-1\right],
        \ear

        \bear \label{sumn}
        s_n(y)=\sum_{l=0}^\infty 
        \frac{(l+1/2)^{2n-1}}{[y^2+(l+1/2)^2]^{n+1/2}},
        \ear
where $n=1,2,3,\dots$ and $y>0$. To transform them we use the
Abel-Plana method of summarizing of convergent series \cite{Evgrafov}. The
essence of this method consists of following: the series is presented as a
contour integral
        \bear
        2i\sum_n f(n+\frac12)=
        2\pi i\sum_n\mathop{\rm res}\limits_{z=n+\frac12} 
        \left[ f(z)\cot\pi(z-\frac12)\right]
        =\int\limits_{\cal C}f(z)\cot\pi(z-\frac12)dz\,,
        \ear
where ${\cal C}$ is a contour which surrounds a region on the complex
plane containing only the poles of $\cot\pi(z-\frac12)$; the poles of
$f(z)$ (if they exist) have to lie outside of the contour ${\cal C}$.
Deforming the contour by a special way and currying out not difficult
algebraic transformations gives the following relation:
        \bear 
        \sum_{l=0}^\infty && f(l+\frac12) 
        = \lim_{\sigma\to0}\left\{
        \int_\sigma^\infty f(x) dx
        \right.
        \left.
        +i\int_{i\sigma}^{i\sigma+\infty}\frac{f(-it)dt}{e^{2\pi t}+1}
        -i\int_{-i\sigma}^{-i\sigma+\infty}\frac{f(it)dt}{e^{2\pi t}+1}
        \right\}.
        \ear
Using this formula we may transform Eqs. (\ref{sum0}) and (\ref{sumn}) as
follows:
        \bear \label{tsum0}
        s_0(y)=&&
        \int_0^\infty\left[\frac{x}{\sqrt{y^2+x^2}}-1\right]dx
        +2\int_0^y\frac{t}{\sqrt{y^2-t^2}}\frac{dt}{e^{2\pi t}+1},
        \ear
        \bear\label{tsumn}
        s_n(y)=&&
        \int_0^\infty\frac{x^{2n-1}dx}{(y^2+x^2)^{n+1/2}}
        -2(-1)^n\lefteqn{~-}\int_0^y\frac{t^{2n-1}}{(y^2-t^2)^{n+1/2}}
        \frac{dt}{e^{2\pi t}+1},
        \ear
where  
        \bear 
        \lefteqn{~-}\int_0^yf(t)dt&=&\lim_{\delta\to 0}
        \left[\int_0^{y-\delta}f(t)dt 
        -  
        \left(
        \begin{array}{l}
        \rm terms~which~diverge \\
        \rm in~the~limit~\delta\to 0
        \end{array}
        \right) \right].
        \nonumber
        \ear
Calculating the first integral in Eq. (\ref{tsum0}) and carring out an
integration by parts in Eq. (\ref{tsumn}) yields 
        \bear \label{trsum0}
        s_0(y)=-y
        +2\int_0^y\frac{t}{\sqrt{y^2-t^2}}\frac{dt}{e^{2\pi t}+1},
        \ear

        \bear \label{trsumn}
        s_n(y)=&&
        -\frac{2}{(2n-1)!!}\int_0^y\frac{dt}{\sqrt{y^2-t^2}}
        \frac{d}{dt}\left(\frac1t\frac{d}{dt}\right)^{n-1}
        \left(\frac{t^{2(n-1)}}{e^{2\pi t}+1}\right).
        \ear

The expressions for the functions $S_n(\epsilon,\mu)$, given by Eqs.
(\ref{s0}) and (\ref{sn}), may be rewritten as
        \bear\label{Sn}
        && S_n(\epsilon,\mu)=\int_0^\infty du 
        \cos ( u\varepsilon ) 
        s_n(\sqrt{u^2+\mu^2}),
        \nonumber 
        \ear
where $\varepsilon=\epsilon r^{-1}f^{1/2}$ and $n=0,1,2,\dots$. 

In order to derive the asymptotical expansion for $S_0(\epsilon,\mu)$ and 
$S_n(\epsilon,\mu)$ in the limit $\epsilon\to 0$ we use the
following asymptotical relations:
        \bear\label{rel1}
        I_1(\varepsilon)\equiv &&\int_0^\infty  
        \sqrt{u^2+\mu^2}\cos(u\varepsilon)du=
        -\frac1{\varepsilon^2}
        -\frac{\mu^2}{2}\left(\ln\frac{\varepsilon\mu}{2}+C\right)
        +\frac{\mu^2}{4}+{\rm O}(\varepsilon^2\ln\varepsilon),
        \ear

        \bear\label{rel2}
        I_2(\varepsilon) & \equiv & \int_0^\infty du \,  
        \cos(u\varepsilon)
        \int_0^{\sqrt{u^2+\mu^2}}\frac{f(t)dt}{\sqrt{u^2+\mu^2-t^2}}
        \nonumber\\
        &=&
        -\left(\ln\frac{\varepsilon}{2}+C\right)
        \int_0^\infty f(t)\,dt
        -\frac12\int_0^\infty f(t)\ln|\mu^2-t^2|\,dt
        +{\rm O}(\varepsilon^2\ln\varepsilon).
        \ear
To prove the first formula (\ref{rel1}) we make the following
transformations: 
        \bear
        I_1(\varepsilon)&=&\int_0^\infty
        \frac{(u^2+\mu^2)\cos(u\varepsilon)}{\sqrt{u^2+\mu^2}}\,du
        \nonumber\\
        &=&\left(-\frac{d^2}{d\varepsilon^2}+\mu^2\right)
        \int_0^\infty \frac{\cos(u\varepsilon)}{\sqrt{u^2+\mu^2}}\,du
        \nonumber\\
        &=&\left(-\frac{d^2}{d\varepsilon^2}+\mu^2\right)
        K_0(\varepsilon\mu)
        \nonumber\\
        &=&-\frac{\mu}{\varepsilon}K_1(\varepsilon\mu).
        \nonumber
        \ear
Here $K_n(y)$ are the Bessel functions of the second kind. Taking into
account the asymptotic of $K_1(y)$ in the limit $y\to 0$:
        $$
        K_1(y)= \frac1y+\frac y2\left(\ln\frac y2+C-\frac12\right)
        +{\rm O}(y^2\ln y),
        $$
we obtain the relation (\ref{rel1}).  

To prove the second formula (\ref{rel2}) we change the order of
integration over $u$ and $t$ in $I_2$ and compute the integral over $u$,
so that 
        \bear
        I_2(\varepsilon)&=&
        \int_0^\mu f(t)\,dt \int_0^\infty 
        \frac{\cos(u\varepsilon)\, du}{\sqrt{u^2+\mu^2-t^2}}
        +\int_\mu^\infty f(t)\,dt \int_{\sqrt{t^2-\mu^2}}^\infty 
        \frac{\cos(u\varepsilon)\, du}{\sqrt{u^2+\mu^2-t^2}}
        \nonumber\\
        &=&\int_0^\mu f(t)K_0\left(\varepsilon\sqrt{\mu^2-t^2}\right)\,dt
        -\frac{\pi}{2}\int_\mu^\infty f(t)
        Y_0\left(\varepsilon\sqrt{t^2-\mu^2}\right)\,dt .
        \ear
Here $K_0(y)$ and $Y_0(y)$ are the Bessel functions of the second kind.
Taking into account the asymptotics of $K_0$ and $Y_0$ in the limit $y\to
0$:
        \bear
        K_0(y)=-(\ln\frac y2+C)+{\rm O}(y^2\ln y),
        \nonumber\\
        Y_0(y)=\frac2\pi(\ln\frac y2+C)+{\rm O}(y^2\ln y),
        \nonumber
        \ear
we can obtain the relation (\ref{rel2}).

Finally, substituting Eqs. (\ref{trsum0}) and (\ref{trsumn}) into Eq.
(\ref{Sn}) and using formulas (\ref{rel1}) and (\ref{rel2}) gives
        \bear \label{appS0}
        S_0(\epsilon,\mu)&=&\frac{r^2}{f\epsilon}
        +\frac12\left(\mu^2-\frac{1}{12}\right)
        \left(\ln\frac{\epsilon\mu}{2}+C\right) 
        -\frac{\mu^2}{2}-\mu^2 S_0(2\pi\mu)+{\rm O}(\epsilon^2\ln\epsilon),
        \ear

        \bear \label{appSn}
        S_n(\epsilon,\mu)&=&
        -\frac{2^{n-1}(n-1)!}{(2n-1)!!}
        \left(\ln\frac{\epsilon\mu}{2}+C\right)
        +\frac{S_n(2\pi\mu)}{(2n-1)!!}+{\rm O}(\epsilon^2\ln\epsilon),
        \ear
where $(2n-1)!!\equiv 1\cdot3\cdot5\cdots(2n-1)$, $n=1,2,3,\dots$ and
$S_0(2\pi\mu)$ and $S_n(2\pi\mu)$ are 
        \beq\label{appS}
        \begin{array}{l}
        \dst
        S_0(2\pi\mu)=\int_0^\infty\frac{x\ln|1-x^2|}
        {e^{2\pi\mu x}+1}dx, \\[12pt]
        \dst
        S_n(2\pi\mu)=\int_0^\infty dx \ln|1-x^2| \frac{d}{dx}
        \left(\frac1x\frac{d}{dx}\right)^{n-1} 
        \frac{x^{2(n-1)}}{e^{2\pi\mu x}+1}.
        \end{array}
        \eeq

\section*{Appendix B: Asymptotics for $S_n(2\pi\mu)$ and $N_n^m(\mu)$} 

\setcounter{equation}{0} 
\renewcommand{\theequation}{B.\arabic{equation}}
In this appendix we obtain asymptotics at large values of argument for the
functions $S_n(2\pi\mu)$ and $N_n^m(\mu)$. Consider the functions
$S_n(2\pi\mu)$ defined by Eqs. (\ref{appS}). Making the substitution
$y=2\pi\mu x$ in Eqs. (\ref{appS}) we find 
        \beq\label{appS_B}
        \begin{array}{l}
        \dst
        S_0(2\pi\mu)=\lambda^2\int_0^\infty
        \frac{y\ln|1-\lambda^2 y^2|}
        {e^{y}+1}\,dy, \\[12pt]
        \dst
        S_n(2\pi\mu)=\int_0^\infty dy \ln|1-\lambda^2 y^2| \,
        \frac{d}{dy}
        \left(\frac1y\frac{d}{dy}\right)^{n-1} 
        \frac{y^{2(n-1)}}{e^{y}+1},
        \end{array}
        \eeq
where $\lambda=(2\pi\mu)^{-1}$ and $\mu=(m^2r^2+\frac14)^{1/2}$. Note that
the integrands in Eqs. (\ref{appS_B}) contain exponential functions and are
exponentially decreasing at large values of $y$. Hence the main contribution
into the integrals is provided with values of integrands in the region $0 < y
< 1$. We are interesting in the case $\mu>\!\!>1$. In this case
$\lambda<\!\!<1$ and $\lambda y <\!\!< 1$ if $0<y<1$. Now we may use the
asymptotical formula $\ln(1-\lambda^2 y^2)= -\lambda^2 y^2-\frac12 \lambda^4
y^4-\frac13 \lambda^6 y^6 +{\rm O}(\lambda^8 y^8)$. Substituting this into
Eqs. (\ref{appS_B}) and integrating gives
        \bear\label{asymS_B}
        &&S_0(2\pi\mu)=-\frac{7}{1920(mr)^4}+{\rm O}((mr)^{-6}), \qquad
        S_1(2\pi\mu)=\frac{1}{24(mr)^2}+{\rm O}((mr)^{-4}), \quad
        \nonumber \\
        &&S_2(2\pi\mu)=-\frac{0.0146}{(mr)^4}+{\rm O}((mr)^{-6}), \qquad
        S_3(2\pi\mu)=\frac{0.03}{(mr)^6}+{\rm O}((mr)^{-8}).
        \ear
Consider now the functions $N_n^m(\mu)$:        
        \bear
        &&N_2^1(\mu)=\sum_{l=0}^\infty
        \frac{l+\frac12}{[\mu^2+(l+\frac12)^2]^2},\quad {\rm and} \quad
        N_3^i(\mu)=\sum_{l=0}^\infty
        \frac{(l+\frac12)^i}{[\mu^2+(l+\frac12)^2]^3},
        \ear
where $i=1,3$. Using the formula (5.1.26.23) from Ref. \cite{PBM} we find
        \bear
        &&N_2^1(\mu)=-\frac{i}{4\mu}\left[\psi'\left(\frac12-i\mu\right)
        -\psi'\left(\frac12+i\mu\right)\right],
        \ear
where $\psi(z)=\Gamma'(z)/\Gamma(z)$ is the digamma function.
It is also easily to see that 
        $$
        N_3^1(\mu)=-\frac{1}{4\mu} \frac{dN_2^1(\mu)}{d\mu}
        \quad {\rm and} \quad N_3^3(\mu)=N_2^1(\mu)-\mu^2
        N_3^1(\mu).
        $$  
Using the asymptotical properties of the digamma function (see, for example,
Ref. \cite{Abram}) we can obtain for large values of $\mu$ the following
asymptotics: 
        \bear
        &&N_2^1(\mu)=\frac{1}{2(mr)^2}+{\rm O}((mr)^{-4}), \qquad
        N_3^1(\mu)=\frac{1}{4(mr)^4}+{\rm O}((mr)^{-6}), \quad
        \nonumber \\
        &&N_3^3(\mu)=\frac{1}{4(mr)^2}+{\rm O}((mr)^{-4}).
        \ear

\newpage
\begin{figure}[htb]
\epsfysize=5cm\epsffile{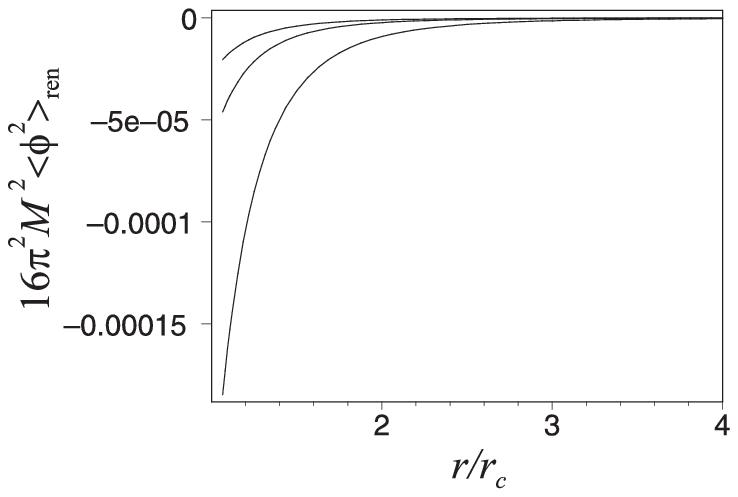}

\caption{The curves in this figure display the value of $\phisq$ for a
massive scalar field in the Schwarzschild spacetime. From bottom to top
the curves correspond to the values $\Lambda=\rg/\rc=10,20,30$.}

\end{figure}

\newpage
\begin{figure}[htb]
\epsfysize=5cm\epsffile{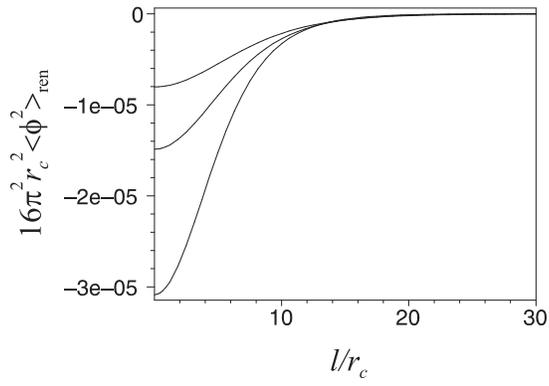}

\caption{The curves in this figure display the value of $\phisq$ for a
massive scalar field in the wormhole spacetime. From bottom to top the
curves correspond to the values $\rho_0=r_0/\rc=10,12,14$.}

\end{figure}

\end{document}